\documentclass{kapproc} 

\usepackage{procps} 
\usepackage[dvips]{graphicx}

\upperandlowercase

\let\footnote\savefootnote

\kluwerbib 

\def\etal{{et~al. }}
\def\ie{{\it i.e. }}

\def\deg{$^{\circ}$}

\def\microKCMB{$\mu\mathrm K_{CMB}$}

\newcommand{\Archeops}{{\sc Archeops }}
\newcommand{\Boomerang}{{\sc Boomerang }}
\newcommand{\Cobe}{{\sc Cobe }}

\newcommand{\Dasi}{{\sc Dasi }}

\newcommand{\Maxima}{{\sc Maxima }}
\newcommand{\Planck}{{\sc Planck-Hfi }}

\newcommand{\OmL}{\Omega_{\Lambda}}

\newcommand{\OmT}{\Omega_{\rm tot}}

\newcommand{\Ombhh}{\Omega_{\rm b}h^2}
\def\ApJ{Astrophys. J.}
\def\ApJL{Astrophys. J. Lett.}
\def\AA{A\&A}

\def\MNRAS{Month. Not. Roy. Ast. Soc.}

\begin{document}

\articletitle[Archeops: an instrument for present and future
cosmology]{Archeops: an instrument for present and future cosmology}


\author{Matthieu Tristram}
\affil{LPSC, Grenoble, France}
\email{tristram@lpsc.in2p3.fr}

\author{on behalf of the \uppercase{A}rcheops \uppercase{C}ollaboration }
\affil{includes scientists from Caltech (USA), Cardiff Univ., CESR
  (Toulouse), CSNSM (Orsay), CRTBT (Grenoble), IAS (Orsay), IAP
  (Paris), IROE (Firenze, Italy), ISN (Grenoble), JPL (USA), LAL
  (Orsay), LAOG (Grenoble), Landau Institute (Russia), La Sapienza
  Univ.  (Roma, Italy), LAOMP (Toulouse), Maynooth Univ. (Ireland),
  Minnesota Univ.  (USA), PCC (Paris), SPP (Saclay)}

\begin{abstract}
  \Archeops is a balloon-borne instrument dedicated to measure the
  cosmic microwave background (CMB) temperature anisotropies. It has,
  in the millimetre domain (from 143 to 545~GHz), a high angular
  resolution (about 10~arcminutes) in order to constrain high $\ell$
  multipoles, as well as a large sky coverage fraction (30\%) in order
  to minimize the cosmic variance.  It has linked, before {\sc WMAP},
  \Cobe large angular scales to the first acoustic peak region.  From
  its results, inflation motivated cosmologies are reinforced with a
  flat Universe ($\OmT=1$ within 3~\%). The dark energy density and
  the baryonic density are in very good agreement with other
  independent estimations based on supernovae measurements and big bang
  nucleosynthesis.  Important results on galactic dust emission
  polarization and their implications for \Planck are also addressed.
\end{abstract}



\section*{Introduction}
\Archeops is a CMB bolometer-based instrument using \Planck technology
that fills a niche where previous experiments were unable to provide
strong constraints.  Namely, \Archeops seeks to join the gap in $\ell$
between the large angular scales as measured by {\sc Cobe}/{\sc Dmr}
and degree-scale experiments, typically for $\ell$ between 10 and 200.
For that purpose, a large sky coverage is needed.  The solution was to
adopt a spinning payload mostly above the atmosphere, scanning the sky
in circles with an elevation of around 41~degrees. The gondola, at a
float altitude above 32~km, spins across the sky at a rate of 2~rpm
which, combined with the Earth rotation, produces well sampled sky map
at 143, 217, 353 and 545~GHz.

\section*{Description of the instrument}
The instrument was designed by adapting concepts put forward for
\Planck and using balloon-borne constraints~(\cite{trapani}) : namely,
an open $^3$He-$^4$He dilution cryostat cooling spiderweb-type
bolometers at 100~mK, cold individual optics with horns at different
temperature stages (0.1, 1.6, 10~K) and an off-axis Gregorian
telescope.  The CMB signal is measured by the 143 and 217~GHz
detectors while interstellar dust emission and atmospheric emission
are monitored with the 353 (polarized) and 545~GHz detectors. The
whole instrument is baffled so as to avoid stray radiation from the
Earth and the balloon.  We report on the first results obtained from
the last flight (12.5 night hours) that was performed from Kiruna
(Sweden) to Russia in February 2002.

\section*{Results}
After being calibrated with the CMB dipole -- in agreement with the
FI\-RAS Galaxy or Jupiter emission -- eight detectors (yielding
effective beams of typically 12~arcminute FWHM) at 143 and 217~GHz are
found to have a sensitivity better than 200~\microKCMB$.s^{1/2}$.
A large part of the data reduction was devoted to removing systematic
effects coming from temperature variations on the various thermal
stages and atmospheric effects.

\begin{figure}
  \centering
  \includegraphics[width=0.65\hsize]{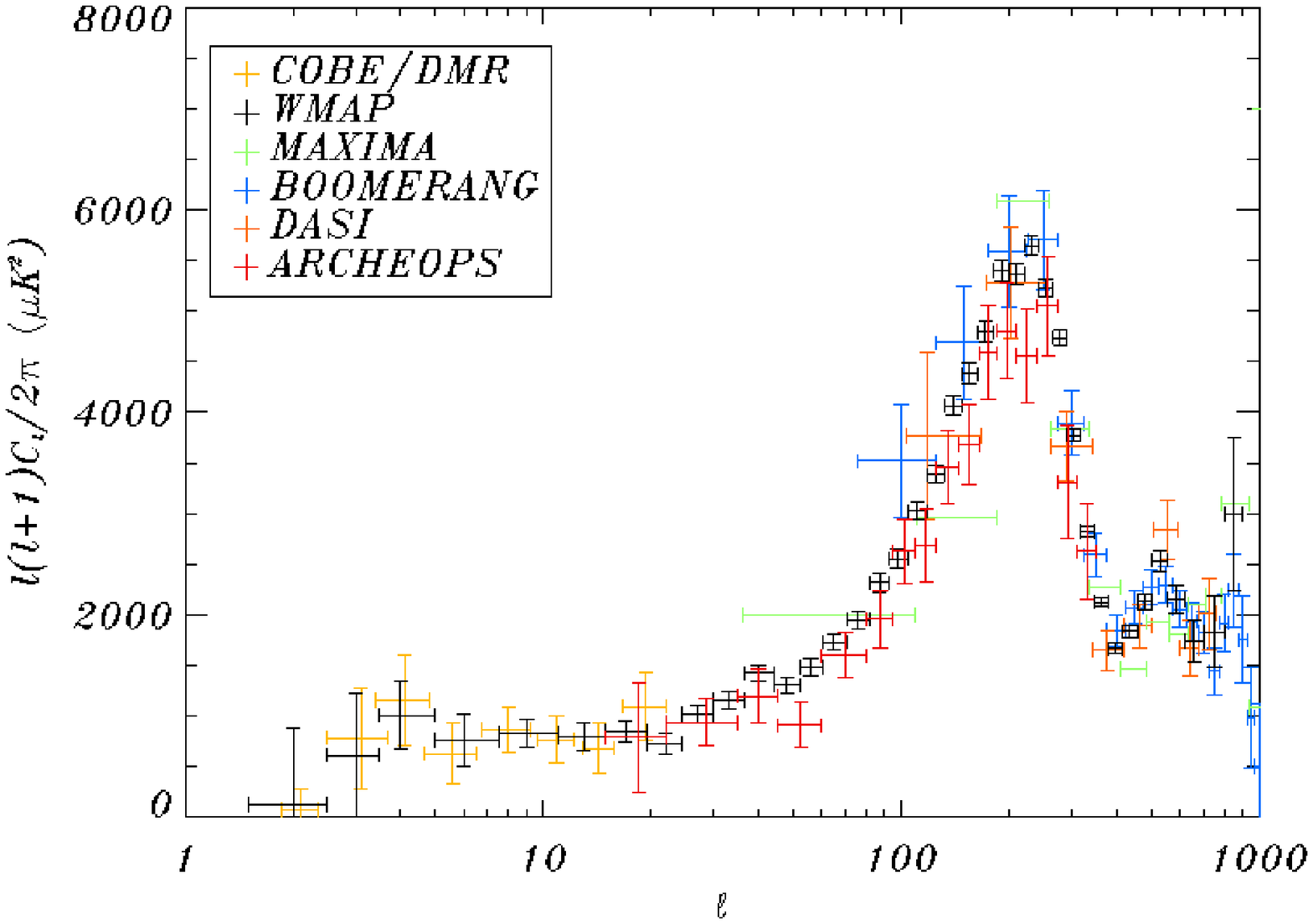}
      \caption{
        \Archeops CMB power spectrum (\cite{benoit_cl}) in 16 bins
        along with other recent experiments \Cobe (\cite{tegmark}),
        {\sc WMAP} (\cite{wmap}), \Maxima (\cite{maxima2}), \Boomerang
        (\cite{boom2}) and \Dasi (\cite{halverson}).}
       \label{figure_cl}
\end{figure}

\cite{benoit_cl} and \cite{benoit_params} show the results of a first
analysis of the data, which are summarized below.  Only the best
bolometer of each CMB channel (143 and 217~GHz) was used.  The data
are cleaned and calibrated, and the pointing is reconstructed from the
stellar sensor data.  The sky power spectrum above a galactic latitude
of 30~\deg (free of foreground contamination) is deduced using a
MASTER-like approach (\cite{hivon}). The observed spectrum is shown in
Fig.~\ref{figure_cl} and compared to a selection of other recent
experiments. Much attention was paid to the possible systematic
effects that could affect the results.  At low $\ell$, dust
contamination and at large $\ell$, bolometer time constant and beam
uncertainties are all found to be negligible with respect to
statistical errors. The sample variance at low $\ell$ and the photon
noise at high $\ell$ are found to be a large fraction of the final
\Archeops error bars in Fig.~\ref{figure_cl}.

\begin{figure}[b]
   \centering
   \includegraphics[width=0.65\hsize]{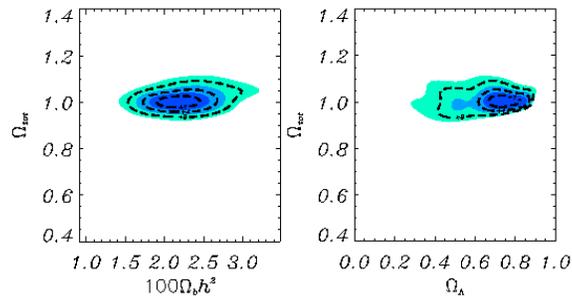}
      \caption{Likelihood contours for 3 of the cosmological parameters:
        total density versus baryonic density and cosmological
        constant. Greyscale corresponds to 2-D limits and dashed line
        to 1-D contours equivalent to 1, 2, and $3\,\sigma$
        thresholds. A Prior on the Hubble constant $H_0=72\pm8
        \rm\,km/s/Mpc$ (68\% CL, \cite{freedman}) has been added.}
       \label{figure_param}
\end{figure}

\section*{Cosmological constraints}
\Archeops provides a precise determination of the first acoustic peak
in terms of position at the multipole $\ell_{\rm peak}=220\pm 6$,
height and width.  Using a large grid of cosmological models with 7
parameters, one can compute their likelihood with respect to the
datasets. An analysis of Archeops data in combination with other CMB
datasets constrains the baryon content of the Universe to a value
$\Ombhh = 0.022^{+0.003}_{-0.004}$ which is compatible with Big-Bang
nucleosynthesis (\cite{omeara}) and with a similar accuracy
(Fig.~\ref{figure_param}).  Using the recent HST determination of the
Hubble constant (\cite{freedman}) leads to tight constraints on the
total density, {\it e.g.}  $\OmT =1.00^{+0.03}_{-0.02}$, \ie the
Universe would be flat.  An excellent absolute calibration consistency
is found between {\sc Cobe}, \Archeops and other CMB experiments
(Fig.~\ref{figure_cl}). All these measurements are fully compatible
with inflation-motivated cosmological models. The constraints shown on
Fig.~\ref{figure_param}~(right), leading to a value of $\OmL =
0.73^{+0.09}_{-0.07}$ for the dark energy content, are independent
from and in good agreement with supernovae measurements
(\cite{perlmutter}) if a flat Universe is assumed.

\section*{Polarized Foregrounds}
The polarized channel at 353GHz gives important results on the
polarization of the emission of the dust in the galactic plane.
Concerning the large scales, we find a diffuse emission polarized at
4-5\% with an orientation mainly perpendicular to the galactic plane.
We found also several clouds of a few square degrees polarized at more
than 10\% in the Gemini and the Cepheus regions. It is interesting to
note that the brightest region, Cygnus, is not polarized. These
results suggest a powerful grain alignment mechanism throughout
interstellar medium.  All the interpretations are developed in
\cite{archeops_foregrounds}.  Interstellar dust polarization emission
will be a major foreground for the detection of the polarized CMB for
{\sc Planck-Hfi}.

\section*{Conclusions}
The measured power spectrum (\cite{benoit_cl}) matches the \Cobe data
and provides for the first time a direct link between the Sachs-Wolfe
plateau and the first acoustic peak.  The measured spectrum is in good
agreement with that predicted by inflation models producing scale-free
adiabatic peturbations and a flat Universe. Finally note that these
results were obtained with only half a day of data.

Use of all available bolometers and of a larger sky fraction should
yield an even more accurate and broader CMB power spectrum in the near
future. The large experience gained on this balloon-borne experiment
is providing a large feedback to the \Planck data processing
community.



\end{document}